\documentclass[sigconf]{acmart}
\AtBeginDocument{%
  }

\setcopyright{none}
\copyrightyear{2026}
\acmYear{2026}
\acmDOI{}
\acmISBN{}
\acmConference{CHI Workshop on Everyday Wearable for Personalized Health and Well-Being, April 13--17, 2026, Barcelona, Spain}

\begin{document}

%%
%% The "title" command has an optional parameter,
%% allowing the author to define a "short title" to be used in page headers.
\title[HeartbeatCam]{HeartbeatCam: Self-Triggered Photo Elicitation of Stress Events Using Wearable Sensing}

%%
%% The "author" command and its associated commands are used to define
%% the authors and their affiliations.
%% Of note is the shared affiliation of the first two authors, and the
%% "authornote" and "authornotemark" commands
%% used to denote shared contribution to the research.
\author{Boyang Zhou}
% \authornote{Both authors contributed equally to this research.}
\orcid{1234-5678-9012}
\affiliation{%
  \institution{University of Washington}
  \city{Seattle}
  \state{WA}
  \country{USA}
}
\email{zby2003@cs.washington.edu}

\author{Zara Dana}
\affiliation{%
  \institution{Supportiv Inc.}
  \city{Seattle}
  \state{WA}
  \country{USA}
}

%%
%% By default, the full list of authors will be used in the page
%% headers. Often, this list is too long, and will overlap
%% other information printed in the page headers. This command allows
%% the author to define a more concise list
%% of authors' names for this purpose.
\renewcommand{\shortauthors}{Zhou et al.}

%%
%% The abstract is a short summary of the work to be presented in the
%% article.
\begin{abstract}
People often recognize what triggered their stress only after the moment has passed.
In therapy, this can become a recurring problem: clients are asked to remember what happened between sessions, but the details that matter (where they were, what they saw and heard, what was happening around them) are easy to lose. We introduce HeartbeatCam, a wearable sensing system that gathers contextual information during moments of elevated stress.
It uses a consumer smartwatch stress signal to trigger capture from an open-source AR glasses camera, recording a sparse image--audio clip that can later be reviewed and annotated.
The system adopts an actionable sensing approach to mental healthcare, using physiological signals along with contextual capture to support collaborative interpretation of stress-triggering moments with mental health professionals.
\end{abstract}

%%
%% The code below is generated by the tool at http://dl.acm.org/ccs.cfm.
%% Please copy and paste the code instead of the example below.
%%


%%
%% Keywords. The author(s) should pick words that accurately describe
%% the work being presented. Separate the keywords with commas.
%% A "teaser" image appears between the author and affiliation
%% information and the body of the document, and typically spans the
%% page.
\begin{teaserfigure}
  \includegraphics[width=\textwidth]{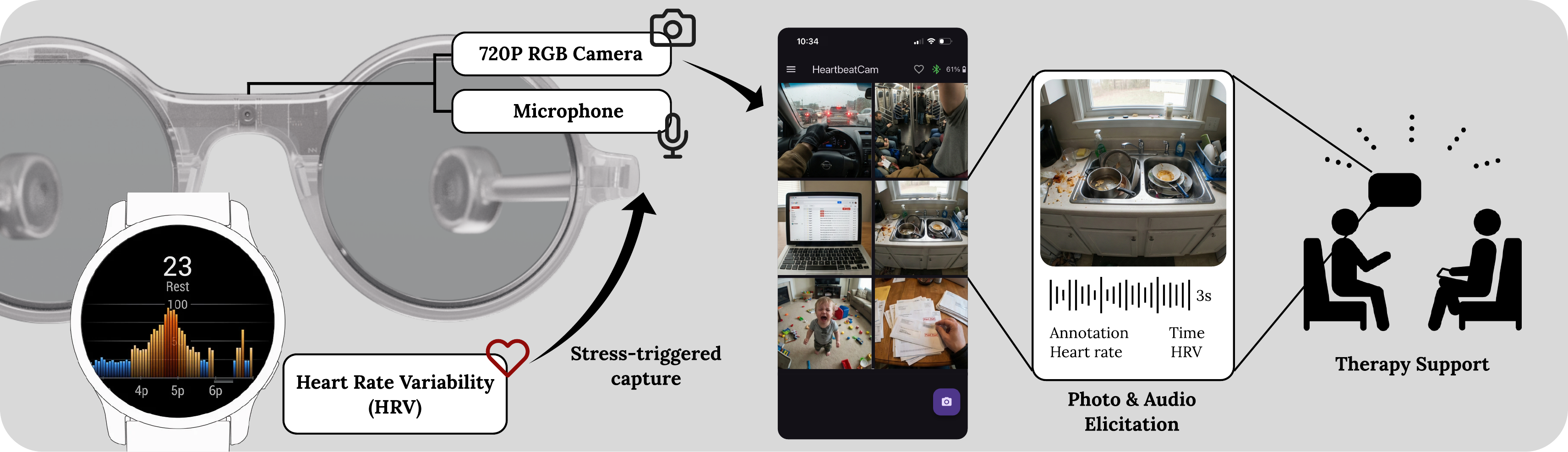}
  \caption{HeartbeatCam automatically captures first-person photos along with audio clips when it detects elevated stress, producing a sparse visual record that clients and therapists can review together during sessions.}
  \Description{System overview of HeartbeatCam. A Garmin smartwatch continuously monitors Heart Rate Variability (HRV); when stress exceeds a threshold, it triggers a stress-triggered capture command to Frame AR glasses, which record a 720p image and 3-second audio clip via the built-in camera and microphone. Captured image--audio pairs are stored and displayed in a companion phone app, where users can annotate each capture with contextual notes alongside heart rate and HRV readings. The resulting records support photo and audio elicitation during therapy sessions.}
  \label{fig:teaser}
\end{teaserfigure}

% \received{todo}
% \received[revised]{todo}
% \received[accepted]{todo}

%%
%% This command processes the author and affiliation and title
%% information and builds the first part of the formatted document.
\maketitle

\section{Introduction}

Many mental health interventions rely on clients noticing patterns in their daily lives, such as what set off a stressful moment and where it happened.
In practice, these details are hard to capture~\cite{lauckCanYouPicture2021}.
When stress is high, opening an app or notebook is rarely the first thing someone is able to do; days later, recall is often incomplete.

Recent work in ubiquitous computing shows how, in clinical settings, passive sensing should move ``beyond detection'' and instead support actions and interpretation in context~\cite{adlerDetectionActionableSensing2024}.
Motivated by this, we introduce \textbf{HeartbeatCam}, a photo elicitation tool that automatically captures first-person context (a visual snapshot and brief audio) when a wrist-worn heart rate sensor detects elevated stress.
The captured image--audio pairs can then be reviewed and annotated by clients and discussed with their therapists to surface stress triggers that would otherwise be difficult to recall.

\textbf{Contributions.} We present (1) the HeartbeatCam system design and prototype, (2) a review-and-annotation workflow intended to fit into therapy conversations, and (3) preliminary stakeholder feedback and open questions around stress sensing and between-session photo elicitation using wearable devices.

\section{Related Work}
\textbf{Photo elicitation.} Photo elicitation methods help people communicate lived experience through visuals and can surface details that are difficult to verbalize alone~\cite{lauckCanYouPicture2021,sinkoPhotoexperiencingReflectiveListening2021,nourseNowYouSee2022}.
Prior works recommend safeguards such as participant-controlled deletion, explicit consent practices, and data minimization in sensitive time/locations, which we incorporated into our system design~\cite{meyerUsingWearableCameras2022, kellyEthicalFrameworkAutomated2013}.

\noindent \textbf{Wearable cameras for recall.}
Wearable cameras such as SenseCam~\cite{hodgesSenseCamWearableCamera2011} capture first-person traces that can cue autobiographical recall.
Images can support richer recall during later reflection, but the benefit depends on whether the capture and review workflow is simple enough to be sustained in everyday life~\cite{alleWearableCamerasAre2017,mairUsingWearableCamera2021}.
Common capture approaches typically rely on manual triggers or fixed-interval recording, which lack the ability to provide just-in-time captures.
% Our work addresses this by using physiological signals to trigger capture only during relevant moments, similar to method established in PulseCam~\cite{niforatosPulseCamBiophysicallyDriven2015}.
Prior explorations of physiologically-driven capture, such as PulseCam~\cite{niforatosPulseCamBiophysicallyDriven2015} and AffectCam~\cite{sasAffectCamArousalAugmented2013}, demonstrated the promise of using heart rate and galvanic skin response to surface moments worth remembering.
We extend this direction to heart rate variability (HRV), a richer cardiac signal that is also readily accessible on consumer wearables.

\noindent \textbf{Sensing for therapy support.} Physiological and behavioral sensing has been used to support therapy processes~\cite{evansUsingSensorCapturedPatientGenerated2024, backEnhancingProlongedExposure2022,saraiyaTechnologyenhancedVivoExposures2022}.
Yet sensed signals are rarely self-explanatory, and clinicians often need context to interpret them~\cite{adlerDetectionActionableSensing2024}.
Our work bridges this gap by attaching visual audio evidence to physiological spikes, transforming abstract data into actionable context.

\noindent \textbf{Measuring stress via HRV.} Heart rate variability (HRV) is widely established as a physiological biomarker for monitoring stress \cite{1996HeartRVA, Velmovitsky2022UsingAWA, Kim2018StressAHA, immanuelHeartRateVariability2023, bessonAssessingClinicalReliability2025}. 
This work uses low-cost consumer-grade devices to infer real-time stress levels from ultra-short HRV (less than 5 minutes).

\section{HeartbeatCam}

We developed HeartbeatCam (\autoref{fig:teaser}), an egocentric photo elicitation system that combines effortless first-person image capture with passive physiological sensing of mental states. 

We utilized open-source AR glasses (Frame, Brilliant Labs) as the primary egocentric image capture device (\autoref{fig:teaser}). The glasses are equipped with a 720p camera and microphone located in the nose bridge. They use a system-on-chip (SoC) design and run on a lightweight Lua-based embedded operating system.

A consumer-grade wrist-worn heart rate sensing device (e.g., Garmin watch) was used to continuously monitor ultra-short heart rate variability (HRV) as a proxy for real-time stress.
Similar to prior work~\cite{Stephenson2021ApplyingHR}, our system first establishes a personalized baseline using one week of continuous HRV measurements. We classify the user as being in a stressful state when the Root Mean Square of Successive Differences (RMSSD) of HRV, calculated over a 25-second sliding window, drops more than 1.5 standard deviations below this baseline~\cite{Stephenson2021ApplyingHR, 2013shortecg, Castaldo2019UltrashortTH}.
Alternatively, third-party stress level APIs (e.g., Garmin Health API~\cite{Garmin}) may be used.

Both the AR glasses and the heart rate sensor connect to a mobile phone via Bluetooth Low Energy (BLE), where data processing and storage take place. When the phone detects elevated stress levels exceeding a predefined threshold, it sends a capture command to the AR glasses. The glasses then take a snapshot and 3 seconds of audio and transmit the raw image back to the phone for storage. We set the maximum frequency of capture to 1 capture/minute to reduce over-sampling during prolonged periods of high tension. The capture session can also be manually paused by double-tapping on the AR glasses.

The mobile interface allows users to annotate each captured image-audio pair with notes and reveals the corresponding image/audio one day after the time of capture, with only metadata available at the time of capture, to avoid overstimulation. The resulting photos can be batch-exported for subsequent review by therapists during sessions.

The resulting collection of captured image–audio pairs provides a structured representation of environmental contexts associated with heightened stress. These records can support therapists in identifying and interpreting visual and auditory cues linked to clients’ stress responses, facilitating more concrete recall and discussion of triggering situations during therapy sessions.

\section{Preliminary Stakeholder Feedback}

We conducted expert walkthroughs of the HeartbeatCam prototype with two mental healthcare professionals (P1, P2). Both experts viewed the system as a promising clinical tool. P2 highlighted its potential to bridge "adherence and compliance gaps"--where clients do not complete between-session assignments--and noted that they are often unaware of how their bodies react to emotions. P1 emphasized that HeartbeatCam could help clients and therapists not only identify stress triggers, but also visualize the client's de-escalation process over time, providing data to gauge objective clinical effectiveness. 

Both experts shared concerns regarding privacy and the therapeutic relationship. P2 mentioned that unforeseen wireless connection issues might induce shame or guilt during highly stressful periods. Furthermore, P2 warned of a potential power imbalance where device logs could be used "against" the client. If providers trust sensor data over a client's self-report (e.g., annotation), it risks damaging therapeutic rapport. To mitigate these risks, P2 recommended that future iterations of HeartbeatCam integrate clinically validated psychology worksheets and stress-coping templates into the self-annotation process. This would promote a more consistent format for self-report, reducing the risk of misinterpretation when sensor data and self-report diverge.

\section{Discussion \& Future Work}

HeartbeatCam demonstrates how sensing technologies can support mental health therapy sessions by producing contextual artifacts that invite discussion.
In line with actionable sensing~\cite{adlerDetectionActionableSensing2024}, our system treats physiology as a prompt to capture context, with meaning constructed through subsequent client-therapist conversation.

A central limitation of the current system is the ambiguity of HRV-based stress inference, and the potential damage to therapeutic rapport when sensor-triggered and self-reported data misalign.
HRV, especially ultra-short HRV, is sensitive to a wide range of physiological states beyond psychological stress, including physical exercise, posture changes, and caffeine intake~\cite{Kim2018StressAHA}.
We plan to further reduce confounds by incorporating additional sensing modalities, such as skin temperature, inertial measurement unit (IMU) data, and geofencing of predefined locations. 

AI-powered image clustering is also an area for future research; vision models can be used to semantically cluster images, enabling a smoother review process and powering an interactive dashboard~\cite{Ji2018InvariantICA}.

Future work should also include field studies with clients who are actively attending therapy, deploying HeartbeatCam over multiple weeks and examining its effectiveness compared to existing between-session worksheets and journaling approaches~\cite{evansUsingSensorCapturedPatientGenerated2024, backEnhancingProlongedExposure2022}.

% \section{Conclusion}
% \TODO{might not need this}
% HeartbeatCam couples consumer stress sensing with sparse egocentric capture to support passive, low-burden photo-elicitation of everyday stress triggers.
% By emphasizing user control and clinician interpretability, it aims to make passive sensing more actionable for mental healthcare.

% \begin{acks}
% To Robert, for the bagels and explaining CMYK and color spaces.
% \end{acks}

%%
%% The next two lines define the bibliography style to be used, and
%% the bibliography file.
\bibliographystyle{ACM-Reference-Format}
\bibliography{references}

%%
%% If your work has an appendix, this is the place to put it.

\end{document}